\documentclass[pre,amsmath, twocolumn, superscriptaddress,showpacs,longbibliography]{revtex4-1}
\usepackage{graphicx}
\usepackage{color}

\begin{document}

\title{Thermodynamics with continuous information flow}

\author{Jordan M. Horowitz}
\affiliation{Department of Physics, University of Massachusetts at Boston, Boston, MA 02125, USA}
\author{Massimiliano Esposito}
\affiliation{Complex Systems and Statistical Mechanics, University of Luxembourg, L-1511 Luxembourg, Luxembourg}

\date{\today}

\begin{abstract}  
We provide a unified thermodynamic formalism describing information transfers in autonomous as well as nonautonomous systems described by stochastic thermodynamics.
We demonstrate how information is continuously generated in an auxiliary system and then transferred to a relevant system that can utilize it to fuel otherwise impossible processes. 
Indeed, while the joint system satisfies the second law, the entropy balance for the relevant system is modified by an information term related to the mutual information rate between the two systems.
We show that many important results previously derived for nonautonomous Maxwell demons can be recovered from our formalism and use a cycle decomposition to analyze the continuous information flow in autonomous systems operating at steady-state. A model system is used to illustrate our findings.
\end{abstract}


\maketitle 

\section{Introduction}

Interacting physical systems not only exchange energy, but also exchange information as they learn about and influence each other.
Harnessing this information flow to do useful tasks is vital in a variety of disciplines: engineers exploit information through feedback to control a system's dynamical evolution~\cite{Astrom,Bechhoefer2005}, biological organisms need to sense their environment in order to adapt~\cite{Smith}, and physicists have been fascinated by the conceptual problems posed by Maxwell's demon for over 100 years~\cite{Leff,Maruyama2009}.

Unfortunately, we are often resigned to qualitative or intuitive descriptions of how information flow influences a system's thermodynamics or energetics, lacking a comprehensive quantitative framework.
Take for example the centrifugal governor whose task it is to continuously monitor the velocity of a motor and to adjust the input of fuel to maintain a constant output power~\cite{Astrom}.
Intuitively, it seems the governor is continuously gathering information about the engine, while simultaneously feeding back that information to control it.
However, the continuous coupling of the input and output makes tweezing apart the measurement from the feedback difficult; so how do we quantify the information in this instance? What is its influence on the governor's thermodynamics?
A similar vagueness occurs in biological sensory adaption, where an organism continuously monitors its environment, while simultaneously changing in response~\cite{Lan2012}.
These  examples typify the difficulties that arise when considering information flow in systems with autonomous dynamics, ones that run continuously on a steady supply of energy.

By contrast, current investigations of information in nonautonomous systems -- ones manipulated by an external agent who drives the system by varying macroscopic external parameters -- are significantly less qualitative~\cite{Allahverdyan2008, Sagawa2008, Cao2009, Jacobs2009, Suzuki2009, Suzuki2010, Sagawa2010, Ponmurugan2010, Sagawa2011b, Horowitz2010, Toyabe2010, Horowitz2011, Horowitz2011b, Granger2011, Esposito2011, Abreu2011, Abreu2012, Bauer2012, Sagawa2012, Still2012, Horowitz2013b, Sagawa2013b, Tasaki2013}.
The paradigm for this situation was established sometime ago by Bennett~\cite{Bennett1982b}, Landauer~\cite{Landauer1961}, and Penrose~\cite{Penrose} in their exorcism of Maxwell's demon~\cite{Leff,Maruyama2009}.
Here, one typically has in mind a thermodynamic engine designed to extract work through feedback, a well-known example being the Szilard engine~\cite{Szilard1964}.
Its operation begins with a measurement whose  outcome is recorded in an auxiliary physical system, often called a memory.
That information is then utilized to extract work by applying a measurement-based feedback protocol to the engine.
For such step-by-step nonautonomous protocols, the information-theoretic mutual information has been identified as a quantitative measure of information useful for thermodynamic analysis. 
In particular, it has been shown to quantify both the minimum energetic cost to measure~\cite{Jacobs2009, Sagawa2010, Granger2011}, as well as bound the maximum work extractable by a feedback engine~\cite{Sagawa2008, Horowitz2010, Ponmurugan2010,Suzuki2010}.
Yet, the information is a static state variable.
For autonomous setups where information constantly flows, there is still no simple way to incorporate the mutual information into the thermodynamics.
Previous studies of the thermodynamics of continuous feedback have been based on coarse-graining~\cite{Kim2007, Munakata2012, Esposito2012, Esposito2013, Munakata2013}, or consider alternative notions of information~\cite{Mehta2012, Barato2013, Diana2013b, Sandberg2014}, such as the transfer entropy~\cite{Ito2013, Hartich2014}, but have no simple connection with the nonautonomous setups.

In this paper, we investigate information processing occurring in small systems where noise is unavoidable~\cite{Lan2012, Mehta2012, Esposito2012b}, so that the dynamics are stochastic. 
To do so, we use the powerful framework of stochastic thermodynamics~\cite{Sekimoto, Seifert2007, Seifert2012, Esposito2010b, Esposito2012, QianPR12a, QianPR12b}, which has been successfully applied to study the nonequilibrium thermodynamics of a diversity of systems, such as (bio)chemical reaction networks, mesoscopic quantum devices, electric circuits, and colloidal particles.
It has also has been verified experimentally in many of these situations~\cite{Liphardt2002, Collin2005, Ciliberto2010, Ciliberto2013, Kung2012, Saira2012, Pekola2013}, including for information operations~\cite{Toyabe2010, Berut2011}.
Here, we establish a general approach to the thermodynamics of information flow between two interacting systems for both autonomous and nonautonomous dynamics.
Not only does our approach naturally incorporate earlier results on nonautonomous systems, but it also provides new perspectives on the information thermodynamics of autonomous systems.
As a consequence, our formalism provides the tools to quantify the thermodynamic cost for utilizing information in a wide array of devices, including sensors or detectors, information engines, and feedback controllers.
We also introduce a method to determine whether the dominant mechanism mediating the interaction between the two coupled systems is energy exchange or is purely informational.

\section{Setup}

\begin{figure}[htb]
\includegraphics[scale=0.65]{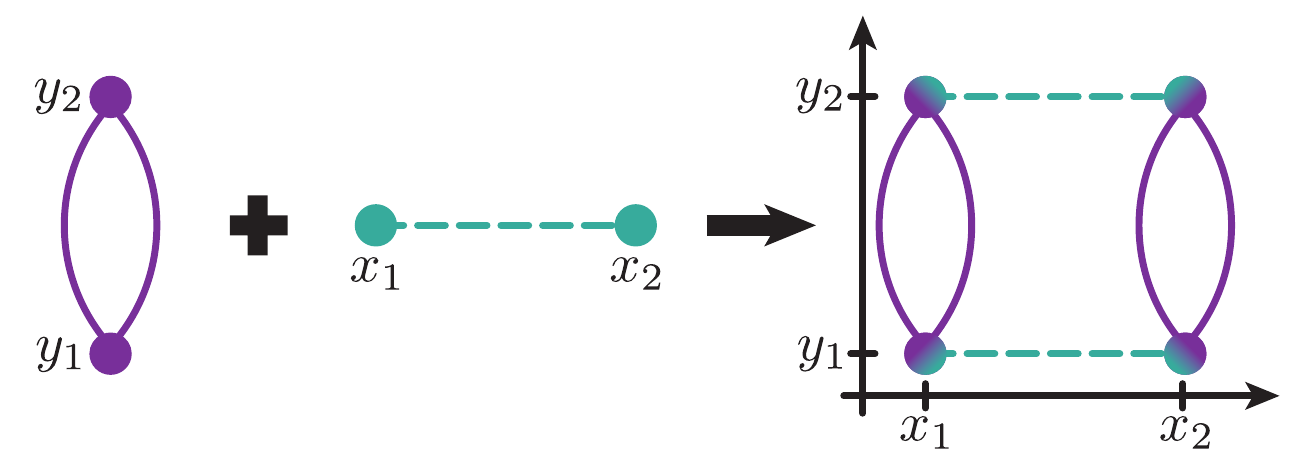}
\caption{Construction of a bipartite graph: a system $X$, with two states $x_1$ and $x_2$ (teal dots) linked by one transition (dashed teal horizontal line), is coupled to an independent system $Y$ that jumps between two states $y_1$ and $y_2$ (purple dots) through two mechanisms (purple vertical lines). The resulting composite system has four states and is bipartite,  since no diagonal edges corresponding to new transition mechanisms are added by the coupling.}
\label{fig:graph}
\end{figure}

\begin{figure*}[htb]
\centering
\includegraphics[scale=.47]{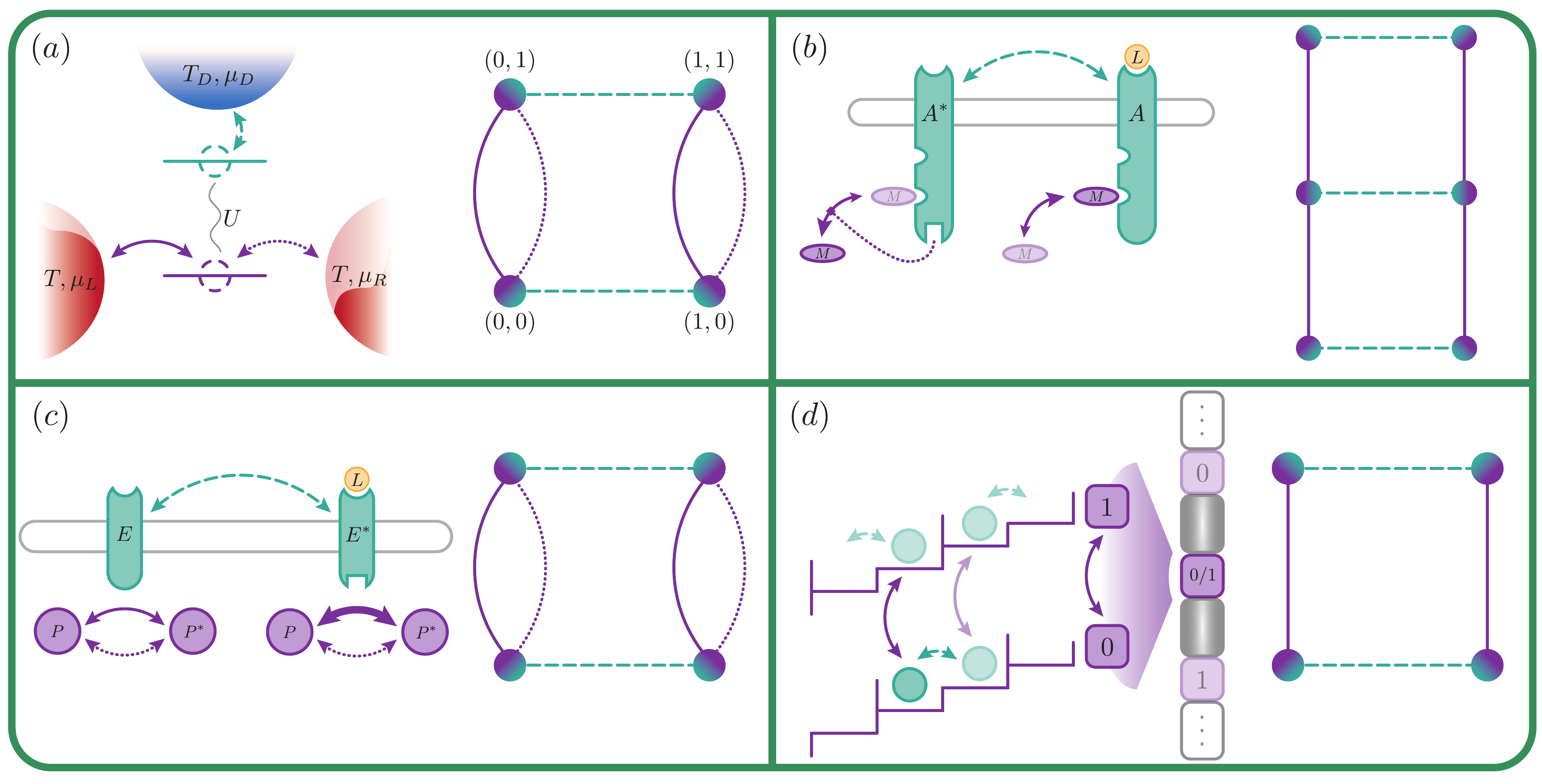}
\caption{
Bipartite examples: 
$(a)$ Quantum dot Maxwell demon: A single-level quantum dot (purple) exchanges electrons with two particle reservoirs (or leads), left and right, at temperature $T$ with chemical potentials $\mu_L$ and $\mu_R$, causing jumps between empty $0$ and filled $1$ states.  An upper ``demon'' dot (teal), connected to a particle reservoir at temperature $T_D$ and chemical potential $\mu_D$, is capacitively  coupled to the lower dot with interaction energy $U$ (gray line).  The ``demon'' dot operates as a measurement and feedback device that can drive a particle current right to left against a chemical potential difference $\Delta \mu=\mu_L-\mu_R>0$~\cite{Esposito2013}.   
$(b)$ Biological sensory adaptation:~a transmembrane enzyme fluctuates between an active configuration $A^*$ and inactive configuration $A$ due to the binding of a extracellular ligand $L$ (yellow).  In the active state $A^*$, the enzyme, through a sequence of reactions (purple dotted), speeds up the removal of bound methyl groups $M$ (purple).  This feedback loop shifts the enzyme's stability, so as to maintain it in the same adapted distribution no matter the ligand concentration~\cite{Morton-Firth1999} 
$(c)$ Biologically inspired model of sensing: a transmembrane receptor enzyme $E$ (teal) is activated by the binding of an extracellular ligand $L$ (yellow). The activated enzyme $E^*$ speeds up one of two nonequilibrium reactions that promote a protein $P$ (purple) to its active configuration $P^*$. In this way, the concentration of $L$ is recorded in the concentration of $P^*$~\cite{Mehta2012,Barato2013}. 
$(d)$ Model information engine: a Brownian particle (teal) diffuses in an tilted energy landscape. One by one a sequence of two-level bits (purple) with states $0$ and $1$ are coupled to the particle. Measurement and feedback is performed through a nonautonomous process that simultaneously flips the bits, while switching the energy landscape. By biasing uphill potential flips with a nonequilibrium force (not shown), the particle can be driven preferentially uphill in order to extract work. The graph depicts the interaction of the particle with one bit and is periodic in the particle's position~\cite{Horowitz2013}.
}
\label{fig:examples}
\end{figure*}

We are interested in coupling together two independent systems $X$ and $Y$, whose discrete states we label $x$ and $y$.
These states, for example, could be the electronic configurations of a quantum dot, or the mechanochemical states of an enzyme.
Each system has its own dynamics dictating the rates at which it makes random transitions among its own states, which we model as Markov processes~\cite{vanKampen}.
It is useful to picture these dynamics occurring on a graph, such as in Fig.~\ref{fig:graph},
where each state is represented by a node (or vertex), and the edges (or links) are the possible transitions. 
Thermodynamics enters by identifying the types of reservoirs -- such as thermal or chemical -- that mediate the transitions along the different edges. 
This requires that the rates describing the transitions satisfy a local detailed balance condition, which allows for a proper identification of the heat exchanged with the reservoirs~\cite{Seifert2012, Esposito2012}. 
We are interested  in combining these two systems into a larger Markovian super-system with states $(x,y)$.
Our rule for coupling the two systems is that we alter the transition rates so the two systems influence each other, but we do not add any new transitions (new links); that is to say we do not fundamentally alter the possible dynamical processes.
Such an arrangement is called \emph{bipartite}~\cite{Barato2013, Barato2013b, Diana2013b, Hartich2014}.
Its key property is that either $X$ jumps or $Y$ jumps, but never both at the same time.
Figure~\ref{fig:examples} illustrates the ubiquity and diversity of this construction with examples drawn from biology, mesoscopic physics, and information thermodynamics.

Since the  total system is Markovian, the time-dependent joint probability distribution $p(x,y)$ evolves according to a master equation 
\begin{equation}\label{eq:masterW}
d_t p(x,y)=\sum_{x^\prime,y^\prime} \bigg( W_{x,x^\prime}^{y,y^\prime}p(x^\prime,y^\prime)-W_{x^\prime,x}^{y^\prime,y}p(x,y) \bigg),
\end{equation}
where $W_{x,x^\prime}^{y,y^\prime}$ is the transition rate at which the system jumps from  $(x^\prime,y^\prime)\to (x,y)$.
The bipartite structure restricts the form of $W$ to
\begin{equation}
W_{x,x^\prime}^{y,y^\prime}=
\left\{\begin{array}{cc}w_{x,x^\prime}^y & x\neq x^\prime; y=y^\prime \\
w_{x}^{y,y^\prime} & x= x^\prime; y\neq y^\prime \\
0 & {\rm otherwise}
\end{array}\right.,
\end{equation}
such that $X$ and $Y$ influence each other's rates, but never jump simultaneously.
In general, $W$ will differ along each link that connects a pair of states, and for nonautonomous processes will depend directly on time; however, we suppress these dependences to keep the notation concise.

Because probability is conserved, it is convenient to recast the master equation as a continuity equation with current  $J_{x,x^\prime}^{y,y^\prime}=W_{x,x^\prime}^{y,y^\prime}p(x^\prime,y^\prime)-W_{x^\prime,x}^{y^\prime,y}p(x,y)$ flowing from $(x^\prime,y^\prime)\to (x,y)$:
\begin{equation}\label{eq:master}
d_t p(x,y)=\sum_{x^\prime,y^\prime}J_{x,x^\prime}^{y,y^\prime}=\sum_{x^\prime} J_{x,x^\prime}^{y}+\sum_{y^\prime}J_{x}^{y,y^\prime},
\end{equation}
where we identified $J_{x,x^\prime}^y=w_{x,x^\prime}^yp(x^\prime,y)-w_{x^\prime,x}^yp(x,y)$ the current from $x^\prime$ to $x$ along $y$, and similarly for $J^{y,y^\prime}_x$.
We see that the bipartite structure allows the current to naturally be divided into two separate flows, one in the $X$-direction and the other in the $Y$-direction.
This is the key property that we exploit in the following.

The joint system is an open system satisfying the second law of thermodynamics~\cite{Esposito2010b,Esposito2011}, which demands that the (irreversible) entropy production always be positive:
\begin{equation}\label{eq:2law}
\begin{split}
\dot S_{\bf i}=d_t S^{XY}+\dot S_{\bf r}\ge 0.
\end{split}
\end{equation}
Here, we use stochastic thermodynamics to identify
\begin{equation}
d_t S^{XY}=\sum_{x\ge x^\prime;y\ge y^\prime}J_{x,x^\prime}^{y,y^\prime}\ln\frac{p(x^\prime, y^\prime)}{p(x,y)}
\end{equation}
as the time derivative of the system's Shannon entropy $S^{XY}=-\sum p(x,y)\ln p(x,y)$, and
\begin{equation}
{\dot S}_{\bf r}=\sum_{x\ge x^\prime; y\ge y^\prime}J_{x,x^\prime}^{y,y^\prime}\ln\frac{W_{x,x^\prime}^{y,y^\prime}}{W_{x^\prime,x}^{y^\prime,y}}
\end{equation}
as the entropy change in the surrounding environment, so that 
\begin{equation}\label{eq:entProd}
{\dot S}_{\bf i}=\sum_{x\ge x^\prime;y\ge y^\prime} J_{x,x^\prime}^{y,y^\prime}\ln\frac{W_{x,x^\prime}^{y,y^\prime}p(x^\prime,y^\prime)}{W_{x^\prime,x}^{y^\prime,y}p(x,y)} \geq 0.
\end{equation}
We have set Boltzmann's constant to unity $k_B=1$, and the over-dot notation, as in ${\dot S}_{\bf r}$, is used to emphasize that such quantities are rates and not the derivative of a function, which is $d_t$.

In general, ${\dot S}_{\bf r}$ quantifies the energy flow to the environment, but its explicit form depends on the types of environmental reservoirs.
For example, when the environment is a single thermal reservoir at temperature $T$, ${\dot S}_{\bf r}$ is proportional to the heat current into the system ${\dot Q}$: ${\dot S}_{\bf r}=-{\dot Q}/T$.
This can be seen by recognizing that local detailed balance requires that $\ln (W_{x,x^\prime}^{y,y^\prime}/W_{x^\prime,x}^{y^\prime,y})=-(\varepsilon_{x,y}-\varepsilon_{x^\prime,y^\prime})/T$ is the change in energy of the system during a jump, which is supplied as heat by the reservoir~\cite{Hill, Seifert2012, Esposito2012}.
Once the reservoirs are identified, the connection to work $W$ and internal energy $U$ can be made through the first law $d_tU={\dot W}+{\dot Q}$.
For one thermal reservoir, $\dot S_{\bf i}=({\dot W}-d_t F)/T$, with $F=U-TS$ the nonequilibrium free energy.
In this way, Eq.~(\ref{eq:2law}) also determines the energetics.

\section{Bipartite thermodynamics and information flow}

Equation~(\ref{eq:2law}) describes the flow of entropy between the system $d_t S^{XY}$ and its environment ${\dot S}_{\bf r}$, but it does not dictate how energy and information flow between the two subsystems.
To make this explicit, observe that each term in Eq.~(\ref{eq:2law}) is a flow, or in other words is a functional of the current.
For any current functional, ${\mathcal A}(J)=\sum J_{x,x^\prime}^{y,y^\prime} A_{x,x^\prime}^{y,y^\prime}$, we can divide it into two contributions just as we split the current in Eq.~(\ref{eq:master}),
\begin{equation}\label{eq:split}
\begin{split}
{\mathcal A}(J)&=\sum_{x\ge x^\prime; y\ge y^\prime} J_{x,x^\prime}^{y} A_{x,x^\prime}^{y,y^\prime}
+\sum_{x\ge x^\prime; y\ge y^\prime} J_{x}^{y,y^\prime} A_{x,x^\prime}^{y,y^\prime} \\
&\equiv {\mathcal A}^X+{\mathcal A}^Y,
\end{split}
\end{equation}
separating the variation in the $X$-direction, ${\mathcal A}^X$, from the $Y$-direction, ${\mathcal A}^Y$.

In the following, we divide the thermodynamics of a bipartite system in this way.
We will find then that we also have to include a new flow, the information flow, defined as the time-variation of the \emph{mutual information} 
\begin{equation}
I=\sum_{x,y} p(x,y)\ln\frac{p(x,y)}{p(x)p(y)}\ge 0,
\end{equation}
which is a measure of correlations that quantifies how much one system ``knows'' about the other.
When $I$ is large, the two systems are highly correlated; whereas small $I$ implies the two systems know little about each other, with $I=0$ signifying the two systems are statistically independent.
Its time derivative is a flow that we divide as $d_t I={\dot I}^X+{\dot I}^Y$, with
\begin{equation}\label{eq:Iflow}
\begin{split}
{\dot I}^X&=\sum_{x\ge x^\prime;y} J_{x,x^\prime}^y \ln \frac{p(y|x)}{p(y|x^\prime)}\\
{\dot I}^Y&=\sum_{x;y\ge y^\prime} J_{x}^{y,y^\prime} \ln \frac{p(x|y)}{p(x|y^\prime)}.
\end{split}
\end{equation}
${\dot I}^X$ and ${\dot I}^Y$ quantify how information sloshes between the two subsystems: when ${\dot I}^X>0$, an $X$ jump on average increases the information $I$. 
In this way, $X$ is learning about or measuring $Y$; vice versa, ${\dot I}^X<0$ signifies that $X$  is decreasing correlations, which can be interpreted as either erasure (information destruction~\cite{Granger2013}) or the consumption of information in order to extract energy, depending on the situation.
Thus, even though the mutual information $I$ is symmetric in $X$ and $Y$, the information flows ${\dot I}^X$ and ${\dot I}^Y$ separately incorporate the $X$ and $Y$ currents, thereby instilling the information flow with an agency and directionally otherwise lacking in the mutual information.

When we apply our separation~(\ref{eq:split}) to the second law~(\ref{eq:2law}), it splits into two positive pieces ${\dot S}_{\bf i}={\dot S}_{\bf i}^X+{\dot S}_{\bf i}^Y$, which are identifiable as the entropy production rates in each subsystem
\begin{equation}\label{eq:2lawSplit}
\begin{split}
\dot S^X_{\bf i}&=d_tS^X+\dot S^X_{\bf r}-\dot I^X\ge 0 \\
\dot S^Y_{\bf i}&=d_tS^Y+\dot S^Y_{\bf r}-\dot I^Y\ge 0.
\end{split}
\end{equation}
Their positivity can be deduced by recognizing the formal similarity of their stochastic thermodynamic representation with Eq.~(\ref{eq:entProd}):
\begin{equation}
\begin{split}
{\dot S}^X_{\bf i}&=\sum_{x\ge x^\prime;y} J_{x,x^\prime}^y \ln \frac{w_{x,x^\prime}^{y}p(x^\prime,y)}{w_{x^\prime,x}^{y}p(x,y)} \ge 0 \\
{\dot S}^Y_{\bf i}&=\sum_{x;y\ge y^\prime} J_{x}^{y,y^\prime} \ln \frac{w^{y,y^\prime}_{x}p(x,y^\prime)}{w^{y^\prime,y}_{x}p(x,y)} \ge 0 .
\end{split}
\end{equation}
Equation (\ref{eq:2lawSplit}) is our first main result. 
It applies to both autonomous and nonautonomous dynamics and quantifies how the entropy balance of $X$ and $Y$ is modified by the flow of information that they exchange.

To gain insight into this separation, imagine for the moment we are not aware of or have access to $Y$; we can only monitor $X$. 
In this case, we still know $S^X$ and can in principle still measure ${\dot S}^X_{\bf r}$ (by monitoring the environmental entropy changes when $X$ jumps along each of its links, averaged over $y$).
Thus, we would assign to $X$ the entropy production rate 
\begin{equation}
\sigma^X=d_t S^X+\dot S^X_{\bf r}=\sum_{x\ge x^\prime;y} J_{x,x^\prime}^y\ln\frac{w_{x,x^\prime}^yp(x^\prime)}{w_{x^\prime,x}^{y}p(x)},
\end{equation}
cf. Eq.~(\ref{eq:2law}).
If $X$ were alone then $\sigma^X\ge 0$, but  the hidden influence of $Y$ allows $\sigma^X<0$: this  seeming violation of the second law is often cited as the signature of a Maxwell demon~\cite{Esposito2012,Ito2013,Sagawa2013b}.
Furthermore, $Y$'s effect on the entropy balance of $X$ in Eq.~(\ref{eq:2lawSplit}) occurs solely through ${\dot I}^X$, which only depends on the sequence of transitions made by $Y$, not on the particular mechanisms driving $Y$.
In other words, from the point of view of $X$, the precise reservoirs in $Y$'s environment are immaterial, and any conclusions regarding $X$ will continue to hold when it is coupled to any system with the same dynamics as $Y$ no matter its environment.
This is especially relevant when $X$ operates as a passive sensory (or detector) for an unknown fluctuating signal~\cite{Mehta2012}.
Here, we typically want to know the energy expended by $X$ to track the signal, but may not be concerned with how that signal is generated.

\section{Nonautonomous Maxwell Demon}

Having introduced our main result, we now explore some of its consequence.
First, we present how known results regarding the thermodynamics of a nonautonomous Maxwell demon emerge~\cite{Bennett1982b,Sagawa2012,Horowitz2013,Sagawa2013b,Tasaki2013}.
In this setup, one system is identified as the engine, say $Y$, and the other is the memory of the demon (or controller), $X$.
The process consists of a sequence of steps (or stages), where either the engine or memory is controlled individually, while the other is held fixed~\footnote{To maintain a system fixed, there are two typical methods.  We can raise large energy barriers (larger than the thermal energy) that confine the system to a region of its state space.  Alternatively, a large time-scale separation will dynamically freeze the slower system.}.
The first step is a measurement where the memory is manipulated so as to form correlations with the engine.
In the subsequent step, that information is used to do a useful task by driving the engine with a feedback protocol that depends on the measurement outcome.

We can apply Eq.~(\ref{eq:2lawSplit}) to each stage of this evolution.
The two subsystems are initially uncorrelated with information $I_{\rm init}=0$. 
During the measurement we drive $X$ with $Y$ fixed to establish an information $I$.
Upon integrating the entropy production (\ref{eq:2lawSplit}) over the course of the interaction interval, we find
\begin{equation}\label{eq:meas}
\begin{split}
\Delta_{\bf i} S_{\rm meas}^X&=\Delta S^X+\Delta_{\bf r} S_{\rm meas}^X-I\ge 0\\
\Delta_{\bf i} S_{\rm meas}^Y&=0,
\end{split}
\end{equation}
where all the information is generated by $X$ (${\dot I}^Y=0$ and $d_tI={\dot I}^X$).
Next, during the feedback step $Y$ evolves with $X$ frozen, and we have
\begin{equation}\label{eq:feed}
\begin{split}
\Delta_{\bf i} S_{\rm fb}^X&= 0 \\
\Delta_{\bf i} S_{\rm fb}^Y&=\Delta S^Y+\Delta_{\bf r} S_{\rm fb}^Y+I\ge 0,
\end{split}
\end{equation}
assuming that all the correlations are consumed, so that the final information is $I_{\rm fin}=0$.
Without the information $I$, the entropy of $Y$ and its environment could only increase, $\Delta S^Y+\Delta_{\bf r}S_{\rm fb}^Y\ge 0$.
However, the information $I$ allows us to circumvent this restriction and reduce the entropy of $Y$ and its environment; in an isothermal process, this could correspond to the conversion of heat into useful work.
In this way, the information $I$ is a resource for $Y$, just like any other source of energy.
As a result, $Y$ is sometimes refereed to as an information engine~\cite{Bauer2012,Horowitz2013}.

Alternatively, we know that $X$ and $Y$ are two subsystems of a composite system whose total entropy production over the two stages of interaction is the sum of Eqs.~(\ref{eq:meas}) and ({\ref{eq:feed}),
\begin{equation}
\Delta_{\bf i} S=\Delta S^Y+\Delta S^X+\Delta_{\bf r} S_{\rm meas}^Y+\Delta_{\bf r} S_{\rm fb}^X\ge 0.
\end{equation}
The information has canceled.
The entropy that was required to establish $I$ during the measurement was the ultimate source of energy that allowed the operation of $Y$.
In other words, by enclosing the memory and engine into one super-system, the information engine reduces to a standard thermodynamic engine.
Thus, within our approach we can recover the accepted resolution of the nonautonomous Maxwell demon paradox.
We can also incorporate within our formalism engines with repeated measurements, in which case each measurement outcome would be recorded in a different subspace of $X$, which is visualized in Fig.~\ref{fig:examples} as a tape of memory cells.
(Expanded discussions of the information flow in these stepwise protocols can be found in Refs.~\cite{Sagawa2012,Sagawa2013b,Horowitz2013}.)

\section{Autonomous information flow}

Equation~(\ref{eq:2lawSplit}) also offers a new perspective on autonomous systems that operate without external driving.
These systems differ as they relax to a time-independent nonequilibrium steady state where constant currents spur continuous energy and information exchange.

In the steady state, the probability distribution is constant, so all time derivatives $d_t$ are zero.
This includes $d_tI=0$, meaning there is only one information flow $\dot{\mathcal I}={\dot I}^X=-{\dot I}^Y$,  and Eq.~(\ref{eq:2lawSplit}) simplifies to
\begin{equation}\label{eq:2lawSplitSS}
\dot{\mathcal S}_{\bf i}^X=\dot{\mathcal S}_{\bf r}^X-\dot{\mathcal I}\ge 0 \qquad \dot{\mathcal S}_{\bf i}^Y=\dot{\mathcal S}_{\bf r}^Y+\dot{\mathcal I}\ge 0,
\end{equation}
where the italics signify time-independent steady state quantities.
An equivalent steady-state expression has been developed independently in Ref.~\cite{Hartich2014} that offers an alternative interpretation for $\dot{\mathcal I}$.

Equation (\ref{eq:2lawSplitSS}) dictates the minimum energetic requirement to continuously process information.
For the sake of discussion, suppose $\dot{\mathcal I}>0$.
In this case, $X$ is operating as a sensor, creating information as it monitors $Y$.
According to Eq.~(\ref{eq:2lawSplitSS}) this task requires that $X$ supply at least that much energy, $\dot{\mathcal S}^X_{\bf r}\ge \dot{\mathcal I}$.
On the other hand, information is being fed into $Y$, where it is a resource that can be used to extract energy, $-\dot{\mathcal S}^Y_{\bf r}\le \dot{\mathcal I}$, either to do work when $Y$ is a feedback engine or maybe to cool a hot reservoir by way of feedback cooling.
Thus, $\dot{\mathcal I}$ bounds the energetic requirements of information processing in autonomous devices, just like in nonautonomous ones.
This observation motivates introducing the thermodynamic efficiencies
\begin{equation}
\varepsilon^X=\frac{\dot{\mathcal I}}{\dot{\mathcal S}^X_{\bf r}}\le 1, \qquad \varepsilon^Y=\frac{|\dot{\mathcal S}^Y_{\bf r}|}{\dot{\mathcal I}}\le 1
\end{equation}
that quantify the effectiveness of the information utilization. 
They refine the traditional efficiency assigned to the super-system when treated as a standard thermodynamic engine: $\varepsilon=|\dot{\mathcal S}^Y_{\bf r}|/\dot{\mathcal S}^X_{\bf r}=\varepsilon^X\varepsilon^Y$.

To further clarify the physical significance of information flow here, recall that the steady-state entropy flow has the simple form $\dot{\mathcal S}_{\bf r}=\sum {\mathcal J}_{x,x^\prime}^{y,y^\prime}{\mathcal F}_{x,x^\prime}^{y,y^\prime}$ of currents ${\mathcal J}_{x,x^\prime}^{y,y^\prime}$ times affinities (or forces) ${\mathcal F}_{x,x^\prime}^{y,y^\prime}=\ln (W_{x,x^\prime}^{y,y^\prime}/W_{x^\prime,x}^{y^\prime,y})$, whose product ${\mathcal J}{\mathcal F}$ gives the rate of energy dissipation into the environment.
The thermodynamic forces -- which depend on the details of the reservoirs -- drive the currents, allowing for the transfer of entropy and energy between different parts of the system.
Comparing with Eq.~(\ref{eq:Iflow}), we see the information flow also has this form,
\begin{equation}\label{eq:Iforce}
\dot{\mathcal I}=\sum_{x\ge x^\prime;y}{\mathcal J}_{x,x^\prime}^y f_{x,x^\prime}^y=-\sum_{x;y\ge y^\prime}{\mathcal J}_x^{y,y^\prime}f_x^{y,y^\prime}.
\end{equation} 
with an information force $f_{x,x^\prime}^y=\ln [p(y|x)/p(y|x^\prime)]$.
In this way, the information acts as a new driving force that can be treated on equal footing with other traditional forces, but is responsible for pushing entropy and energy \emph{between} the two subsystems.

Deeper insight is gained when we take into account the graph structure of the state space.
Sometime ago, Hill~\cite{Hill} and Schnakenberg~\cite{Schnakenberg1976} observed that due to probability conservation in the steady state not all currents are independent (just like Kirchoff's laws for currents in electric circuits).
Only a smaller subset specify the thermodynamics.
These independent currents are those that flow around a \emph{fundamental set of cycles} of the network, like the ones in Fig.~\ref{fig:cycle}. 
Each such cycle is a directed sequence of connected nodes with the same initial and terminal node: ${\mathcal C}=\{(x_1,y_1)\to(x_2,y_2)\cdots\to(x_1,y_1)\}$. 
The fundamental cycles constitute the set of cycles in terms of which all other cycles can be expressed by linear combination. 
This set is not unique and methods to identify them can be found in Refs.~\cite{Hill,Schnakenberg1976, Andrieux2007}.
\begin{figure}[htb]
\centering
\includegraphics[scale=0.5]{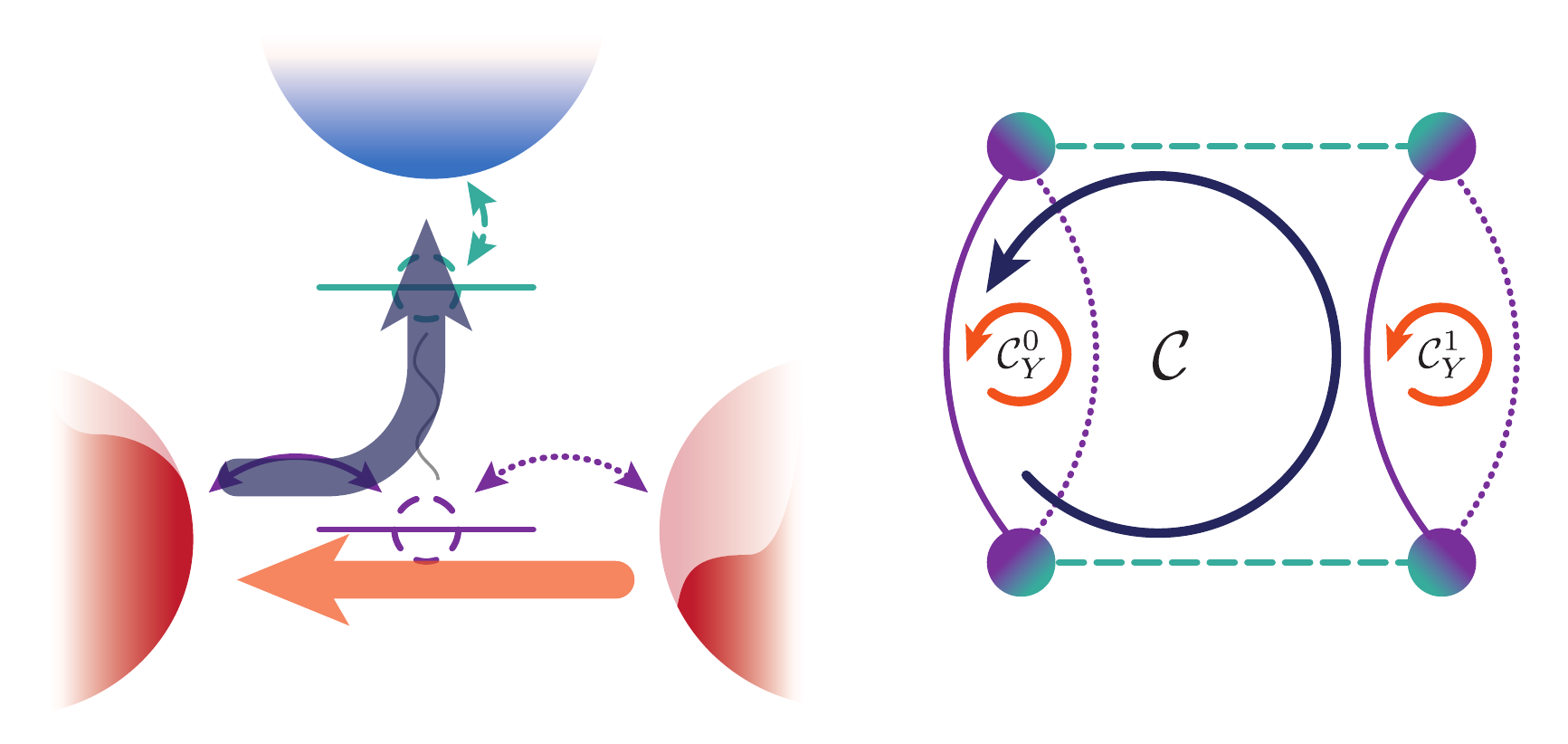}
\caption{Illustration of cycles in a bipartite graph: the global cycle ${\mathcal C}$ (blue arrows) has links in both systems (dashed teal/horizontal and solid purple/vertical), and its current supports an energy flow between the two subsystems. Local cycles of $Y$, ${\mathcal C}^0_Y$ and ${\mathcal C}_Y^1$  (orange arrows), describe particle flow between the two particle reservoirs of the lower system. }
\label{fig:cycle}
\end{figure}
To each fundamental cycle, we assign a current ${\mathcal J}({\mathcal C})$, representing the rate at which probability flows around the cycle, and assign an affinity ${\mathcal F}({\mathcal C})=\sum {\mathcal F}_{x,x^\prime}^{y,y^\prime}$ as the sum of the affinities along the links in ${\mathcal C}$.
It is these cycle currents that capture the mesoscopic fluxes that transfer energy through the system between reservoirs.
The key observation is that the entropy production at steady state can be expressed as ${\dot S}_{\bf r}=\sum_{\mathcal C}{\mathcal J}({\mathcal C}){\mathcal F}({\mathcal C})$ where the sum extends over the fundamental cycles.

In bipartite systems, we can distinguish two types of cycles: global ones and local ones, as in Fig.~\ref{fig:cycle}.
Local cycles are confined to one subsystem, such as for $X$ there is ${\mathcal C}_X=\{(x_1,y)\to(x_2,y)\cdots\to(x_1,y)\}$ (where $y$ is fixed).
They support the internal subsystem flows.
Each local cycle has a local affinity as before, such as ${\mathcal F}^X({\mathcal C}_X)$.
On the other hand, a global cycle $\mathcal{C}$ contains $X$ \emph{and} $Y$ links, so a current flowing around a global cycle pulls with it energy and entropy from one subsystem to the other.
They exclusively result from the coupling of the two subsystems. 
The global affinities ${\mathcal F}({\mathcal C})={\mathcal F}^X({\mathcal C})+{\mathcal F}^Y({\mathcal C})$ affect both systems, so it is useful to split their effect on $X$, ${\mathcal F}^X({\mathcal C})=\sum{\mathcal F}_{x,x^\prime}^y$, from that on $Y$, ${\mathcal F}^Y({\mathcal C})=\sum{\mathcal F}_{x}^{y,y^\prime}$.
Furthermore, the information flow only occurs on global cycles, since it flows \emph{between} the subsystems: $\dot{\mathcal I}=\sum_{\mathcal C} {\mathcal J}({\mathcal C}){\mathcal F}^I({\mathcal C})$, with information affinity ${\mathcal F}^I({\mathcal C})=\sum f_{x,x^\prime}^y=-\sum f^{y,y^\prime}_x$.

Combining these observations on the cycle decomposition, we can rewrite Eq.~(\ref{eq:2lawSplitSS}) to arrive at our second main result (as sketched in Appendix~\ref{sec:cycle})
\begin{equation}\label{eq:cycleEnt}
\begin{split}
\dot{\mathcal S}_{\bf i}^X&=\sum_{\mathcal C}{\mathcal J}(\mathcal C)\left[{\mathcal F}^X(\mathcal C)-{\mathcal F}^I(\mathcal C)\right]+\sum_{{\mathcal C}_X} {\mathcal J}({\mathcal C}_X){\mathcal F}^X({\mathcal C}_X) \\
\dot{\mathcal S}_{\bf i}^Y&=\sum_{\mathcal C}{\mathcal J}({\mathcal C})\left[{\mathcal F}^Y({\mathcal C})+{\mathcal F}^I({\mathcal C})\right]+\sum_{{\mathcal C}_Y}{\mathcal J}({\mathcal C}_Y){\mathcal F}^Y({\mathcal C}_Y).
\end{split}
\end{equation}
This separation of the thermodynamics into global and local cycles is a powerful tool for distinguishing different methods of entropy and energy transfer.
In particular, only flows on the global cycles ${\mathcal C}$ are responsible for direct energy transfer between the subsystems.
For example, in one revolution of ${\mathcal C}$ any energy extracted by ${\mathcal F}^X({\mathcal C})$ will be deposited in $Y$'s environment by ${\mathcal F}^Y({\mathcal C})$.
In the absence of global affinities, ${\mathcal F}^X({\mathcal C})={\mathcal F}^Y({\mathcal C})=0$, energy can only be transferred indirectly by way of an information flow mediated by ${\mathcal F}^I$.
These observations suggest identifying two generic, interaction regimes based on whether the interaction is driven by energy or information.
From the point of view of $X$, when its global affinities are small (${\mathcal F}^X({\mathcal C})\ll\mathcal{F}^I(\mathcal{C})$) the dominant force driving $X$ is information, in which case we say we are in an
\emph{information dominated} regime.
On the other hand, when ${\mathcal F}^X({\mathcal C})\gg{\mathcal F}^I({\mathcal C})$, we are in an \emph{energy dominated} regime where the interaction is powered by energy, not information.
Distinguishing these regimes allows one to identify the driving mechanisms of energy and information transfer, offering a refined understanding of the thermodynamics of information processing.

\section{Example: coupled quantum dots}

To make the above discussion concrete, we now analyze the information thermodynamics of the double quantum dot model pictured in Figs.~\ref{fig:examples}$(a)$ and \ref{fig:cycle}, which has been studied 
extensively both theoretically and experimentally~\cite{BrandesSchallerPRB10,ButtikerSanchez10PRL,ButtikerSanchez12EPL,ButtikerSanchez12EPLerrat,EspositoCuetaraGaspard11,Fujisawa09PRL,Gossard07PRL}.

The device is composed of two single-level quantum dots.
The lower dot in Fig.~\ref{fig:examples}$(a)$, $Y$, exchanges electrons with two leads $\nu=L,R$ at temperature $T$ and chemical potentials $\mu_{\nu}$.
When filled ($y=1$), it has energy $\epsilon_Y$, and when empty ($y=0$), its energy is zero.
In the absence of the second dot, an electronic current flows from left to right down the chemical potential gradient $\Delta \mu=\mu_L-\mu_R> 0$, which we take in to be the negative direction, ${\mathcal J}_e<0$ (opposite the thick orange arrow in Fig.~\ref{fig:cycle}).
The second, upper dot $X$ is connected with a separate lead at a colder temperature $T_D<T$ with chemical potential $\mu_D$. 
It has energy $\epsilon_X$ when filled ($x=1$) and zero when empty ($x=0$).
In absence of the first dot, $X$ will always reach equilibrium with its lead.  
The coupling between the two dots is effected through a capacitive interaction of strength $U$, such that when both dots are filled $(x,y)=(1,1)$ the energy is $\epsilon_X+\epsilon_Y+U$.
The model is finally fixed by setting the rates.
Electron transfers in and out of the upper $X$-dot are given by the rates $W^{y}_{10} = \Gamma f_y$ and $W^y_{01} = \Gamma (1-f_y)$ respectively, where $f_y=(1+\exp{\{(\epsilon_X+y U-\mu_D)/T_D\}})^{-1}$.
The transfers in and out of the lower $Y$-dot have rates $W^{10,(\nu)}_x = \Gamma_x^{(\nu)} f^{(\nu)}_x$ and $W^{01,(\nu)}_{x} = \Gamma_x^{(\nu)} (1-f^{(\nu)}_x)$ where $f^{(\nu)}_x=(1+\exp{\{(\epsilon_Y+x U-\mu_{\nu})/T\}})^{-1}$. 
Notice that we had to specify the lead, left or right, responsible for the transition. 
We have assumed that the density of states of the lead in contact with $X$ is flat, so that $\Gamma$ does not depend on $y$; while the leads in contact with $Y$ have a non-constant density of states, so that $\Gamma_x^{(\nu)}$ depends on the state of the $X$-dot. 

For this model there are three fundamental cycles, depicted in Fig.~\ref{fig:cycle}: one global with current ${\mathcal J}({\mathcal C})$, and two local with currents ${\mathcal J}({\mathcal C}_Y^0)$ and ${\mathcal J}({\mathcal C}_Y^1)$.
The last two currents represent the two contributions to the flow of electrons from the right to left lead of the $Y$-dot denoted ${\mathcal J}_e={\mathcal J}({\mathcal C}_Y^{0})+{\mathcal J}({\mathcal C}_Y^{1})$.
With this decomposition, we can express the total, steady-state entropy production rate~(\ref{eq:2law}) as
\begin{equation}\label{ModelEP}
\dot{\mathcal S}_{\bf i}= -{\mathcal J}_e\frac{\Delta \mu}{T} +  {\mathcal J}({\mathcal C}) \left(\frac{U}{T_D}-\frac{U}{T}\right) \geq 0.
\end{equation}
A more refined picture is offered by Eq.~(\ref{eq:cycleEnt}).
Noting that there is only one global cycle, we have for the information flow $\dot{\mathcal I} = {\mathcal J}({\mathcal C}) {\mathcal F}^I({\mathcal C})$ with information force
\begin{equation}\label{eq:modelFI}
{\mathcal F}^I({\mathcal C})=- \ln \frac{p(x=1|y=1)p(x=0|y=0)}{p(x=1|y=0)p(x=0|y=1)}.
\end{equation}
Then the splitting in Eq.~(\ref{eq:cycleEnt}) reads
\begin{eqnarray}
\dot{\mathcal S}_{\bf i}^X&=& {\mathcal J}({\mathcal C}) \left[\frac{U}{T_D} - {\mathcal F}^I({\mathcal C})\right] \geq 0 \label{ModelEPsplit2} \\
\dot{\mathcal S}_{\bf i}^Y&=& -{\mathcal J}_e\frac{\Delta\mu}{T}  + {\mathcal J}({\mathcal C}) \left[{\mathcal F}^I({\mathcal C})-\frac{U}{T}\right] \geq 0 \label{ModelEPsplit1} .
\end{eqnarray}

\begin{figure}[t]
\centering
\includegraphics[scale=0.55]{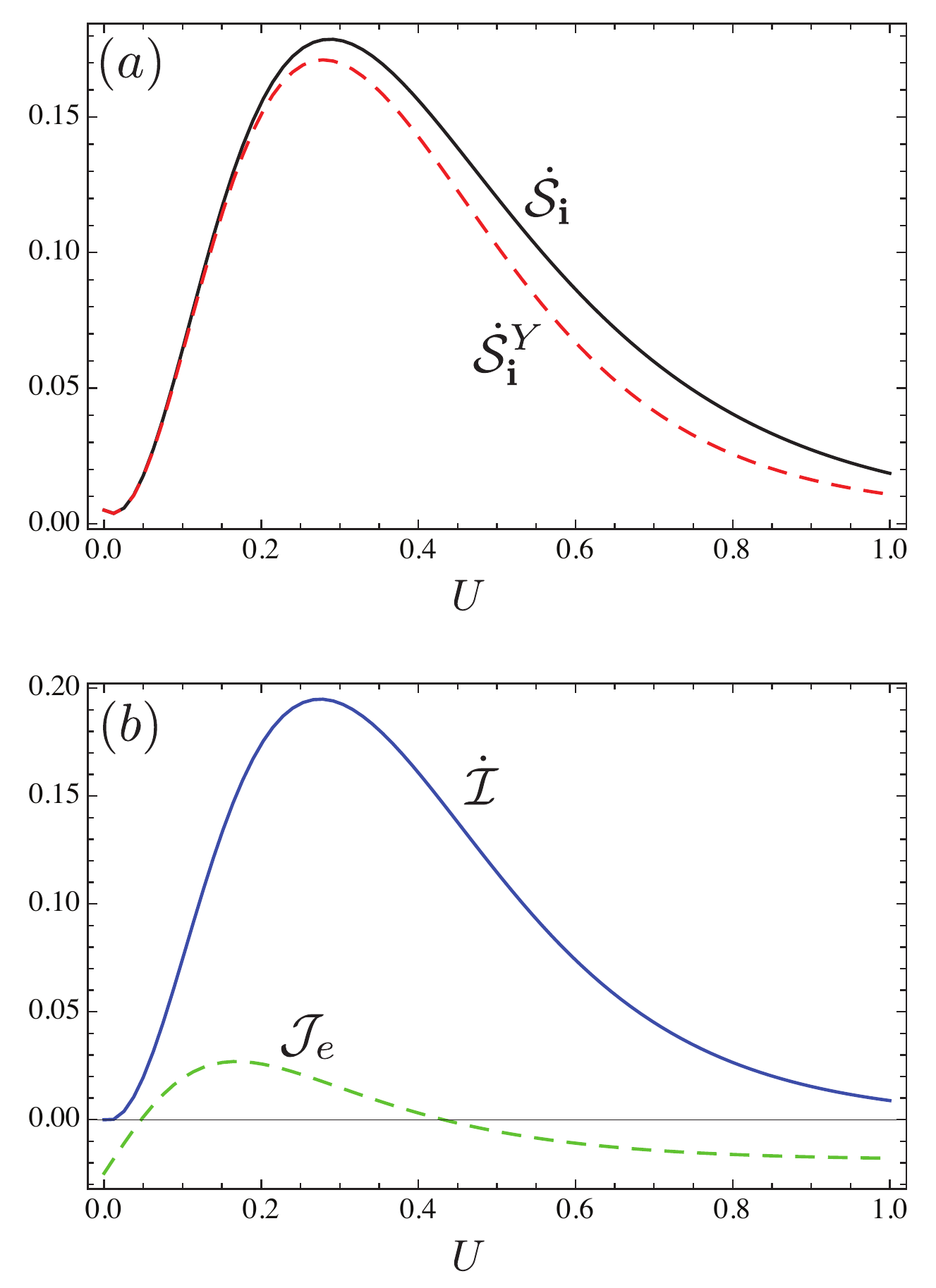}
\caption{Information engine: $(a)$ Plot of total entropy production $\dot{\mathcal S}_{\bf i}$ (black) and entropy production in lower dot $\dot{\mathcal S}^Y_{\bf i}$ (dashed red) as well as $(b)$ information flow $\dot{\mathcal I}$ (blue) and electronic current ${\mathcal J}_e$ (dashed green) as a function of interaction energy $U$. 
Parameters:  $\mu_D=1-U/2$, $\mu_{L}=1.1$, $\mu_{R}=0.9$, $T_D=0.1$, $T=1$, $\epsilon_X=\epsilon_Y=1$, $\Gamma=100$, $\Gamma_{0}^{(L)}=\Gamma_{1}^{(R)}=1.5$ and $\Gamma_{1}^{(L)}=\Gamma_{0}^{(R)}=0.5$.}
\label{fig:DoubleDotModel}
\end{figure}

This device can operate in two modes, either as an information engine or a feedback refrigerator.
The information engine regime occurs when the time-scale of the upper dot $X$ is faster than the lower dot $Y$.
In this limit, $X$ is able to rapidly adapt to the variations in $Y$ allowing it to track and then feedback on $Y$.
Figure~\ref{fig:DoubleDotModel} depicts the thermodynamics in this limit.
We see that there is a regime ($U\approx 0.05 - 0.45$) where the electronic current ${\mathcal J}_e > 0$ is pumped against the bias $\Delta \mu>0$ (Fig.~\ref{fig:DoubleDotModel}$(b)$).
From the global perspective (\ref{ModelEP}), the fuel for this pump is the heat flow $U {\mathcal J}({\mathcal C})>0$ from the hot leads to the cold lead $T_D<T$. 
In other words, the system operates as a thermoelectric device. 
From the information point of view, the only positive term in Eq.~(\ref{ModelEPsplit1}) that can pump the current by compensating the negative $-{\mathcal J}_e\Delta \mu/T<0$ is the information flow $\dot{\mathcal I}={\mathcal J}{\mathcal F}^I>0$ (Fig.~\ref{fig:DoubleDotModel}$(b)$). 
Furthermore, since $X$ is faster than $Y$, the conditional probabilities $p(x|y)$ in Eq.~(\ref{eq:modelFI}) are almost locally equilibrated, which implies ${\mathcal F}^I({\mathcal C})=U/T_D$. 
As a result $\dot{\mathcal S}_{\bf i}^X\approx 0$,  and $\dot{\mathcal S}_{\bf i}\approx \dot{\mathcal S}^Y_i$, as illustrated in Fig.~\ref{fig:DoubleDotModel}$(a)$.
This echoes an observation made in Refs.~\cite{Horowitz2013,Esposito2014} that the most thermodynamically efficient controller is fast enough to instantly equilibrate, so that the measurement and feedback are implemented reversibly.
The ideal Maxwell demon limit of this model (studied in Ref. \cite{Esposito2012}) is $U \to 0$ and $T_D \to 0$ keeping $U/T_D$ finite. 
In this limit, the energetic effects (of order $U$) disappear from Eq.~(\ref{ModelEPsplit1}), and we enter an information dominated regime
\begin{equation}
\dot{\mathcal S}_{\bf i}^Y =- {\mathcal J}_e \frac{\Delta\mu}{T} + \dot{\mathcal I}.
\end{equation}

\begin{figure}[t]
\centering
\includegraphics[scale=0.55]{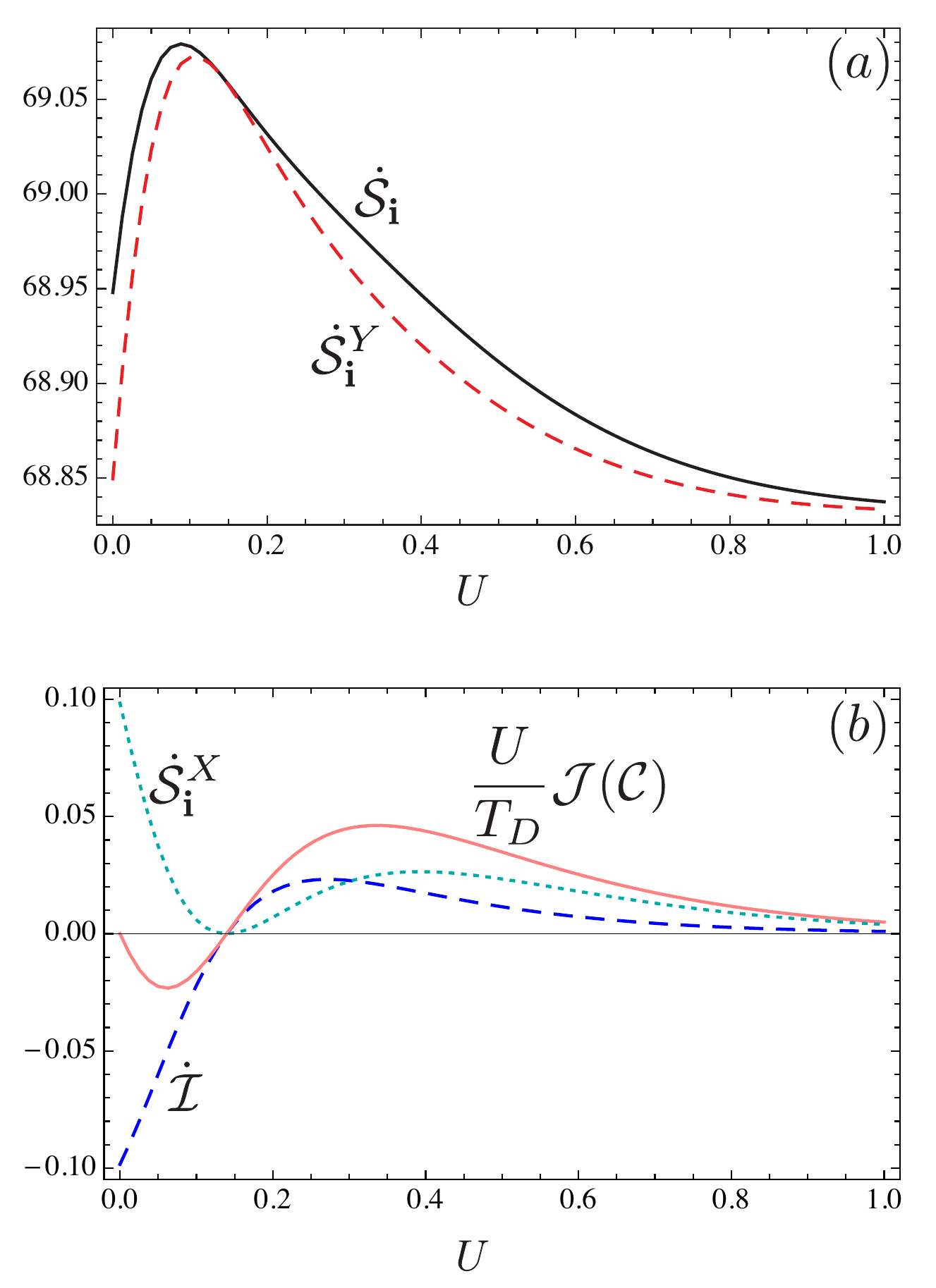}
\caption{Feedback cooling: $(a)$ Plot of total entropy production $\dot{\mathcal S}_{\bf i}$ (black) and entropy production in lower dot $\dot{\mathcal S}^Y_{\bf i}$ (dashed red) as well as $(b)$ information flow $\dot{\mathcal I}$ (dashed blue), entropy production in cooled upper dot, $\dot{\mathcal S}^X_{\bf i}$ (dotted cyan), and energetic current $U{\mathcal J}({\mathcal C})/T_D$ (red) flowing into the reservoir in contact with the upper dot as a function of the interaction energy $U$. 
Parameters: $\mu_D=1-U/2$, $\mu_{L}=0$, $\mu_{R}=3$, $T_D=0.1$, $T=1$, $\epsilon_X=\epsilon_Y=1$, $\Gamma=1$, $\Gamma_{0}^{(L)}=\Gamma_{1}^{(R)}=50$ and $\Gamma_{1}^{(L)}=\Gamma_{0}^{(R)}=150$.}
\label{fig:DoubleDotModel2}
\end{figure}

When the lower dot $Y$ is faster than the upper dot $X$, the model can also operate as a feedback refrigerator that cools the upper dot's reservoir at $T_D<T$.
In Fig.~\ref{fig:DoubleDotModel2}$(b)$, when $U$ is small, the electronic current ${\mathcal J}_e < 0$ flows along the bias $\Delta \mu$ from left to right, cooling the cold lead by extracting heat at a rate $U {\mathcal J}({\mathcal C})<0$. 
From the information perspective, the cooling is fueled by the information $-\dot{\mathcal I}$ provided by the lower dot $Y$. 
This information generation in dot $Y$ is inefficient, since it accounts for the majority of the dissipation, as $\dot{\mathcal S}_{\bf i} \approx \dot{\mathcal S}^Y_{\bf i}$ in Fig.~\ref{fig:DoubleDotModel2}$(a)$. 
However, the information consumption occurring in $X$ to cool is efficient. 
At $U \approx 0.14$, it even reaches equilibrium (i.e. $\dot{\mathcal S}_{\bf i}^X \approx 0$), while the full dissipation $\dot{\mathcal S}_{\bf i}$ remains large.  
At higher values of $U$, the heat flows change direction, and the refrigeration regime is lost,  $U{\mathcal J}({\mathcal C})>0$.

\section{Discussion}

In bipartite systems, information flow describes how two interacting systems learn about and react to each other.
In particular, it bounds the thermodynamics and energetics of each system individually, in this way refining the second law of thermodynamics.
We can view this separation as a type of coarse-graining, where from the point of view of a relevant system, we ignore the transitions of a secondary (or auxiliary) system.
This is a weaker coarse-graining than considered in previous approaches to the thermodynamics of continuous feedback, where the auxiliary system is completely removed (or integrated out) from the description~\cite{Kim2007, Munakata2012, Esposito2012, Esposito2013, Munakata2013}.
It seems that completely removing the auxiliary system is too extreme and ends up removing relevant correlations necessary to establish the connection to earlier results on nonautonomous Maxwell demons.

While we have analyzed the effect of information on the energy flow between reservoirs, we have avoided discussing how one subsystem may do work on the other.
In general, there is no unique way to define such a work, because there is no unique way to partition the total energy between the internal energy of $X$ and internal energy of $Y$~\cite{Jarzynski2006b,Horowitz2007,Campisi2010b,Peliti2008}, though a generic prescription has been proposed~\cite{Boukobza2006}. 
We do contend though that for each physical situation, there is an interesting choice.
Pursuing such an analysis of the work, would be a worthwhile direction for future work.

It is important to note that not all models of Maxwell demons or information engines are bipartite.
One such example was recently devised by Mandal and Jarzynski~\cite{Mandal2012} and has been subsequently adapted and studied in Refs.~\cite{Barato2013c,Mandal2013,Deffner2013,Hoppenau2014}.
These models rectify entropy into work and in this way are considered a Maxwell demon.
However, since they lack a bipartite structure it seems that mutual information does not play a significant role in the thermodynamic analysis.
This illustrates an important point: a low entropy state, such as a memory, is a source of free energy that can be converted into work, but that process need not rely on mutual information as a medium.
Nevertheless, a large class of physically and biologically relevant systems are bipartite.

Finally, the approach we have presented here can be extended beyond classical discrete systems.
Continuous space offers a natural generalization.
Taking the continuous limit of a bipartite master equation (\ref{eq:master}) would result in a bipartite Fokker-Planck equation upon which information flow in diffusive processes could be identified. 
Results obtained in Ref.~\cite{Allahverdyan2009} can be seen as a step in this direction.
This is especially relevant to make comparisons with the literature on optimal stochastic control, which is almost exclusively framed in continuous space~\cite{Astrom,Bechhoefer2005}.
Another interesting extension would be to quantum systems, where the thermodynamics of bipartite systems is already of interest~\cite{Boukobza2006}.
In particular, quantum feedback control can naturally be framed as the interaction of two systems~\cite{Horowitz2013c}, such as in sideband cooling~\cite{Tian2009}.
\section*{Acknowledgments}

J.~M.~H. is supported financially by  ARO MURI grant W911NF-11-1-0268.
M. E. is supported by the National Research Fund, Luxembourg in the frame of project FNR/A11/02. 
 This work was also partially supported by ENFASIS (Spanish Government).

\appendix

\section{Cycle decomposition}\label{sec:cycle}

To obtain the cycle decomposition in Eq.~(\ref{eq:cycleEnt}), we must first identify a fundamental set of oriented cycles.  
A graph-theoretic method to identify this fundamental set can be found in Refs.~\cite{Schnakenberg1976,Andrieux2007}.
Once we have identified this set, for each fundamental cycle ${\mathcal C}$ we define the function $\delta_{x,x^\prime}^{y,y^\prime}({\mathcal C})$ that is $+1$ if the $(x^\prime,y^\prime)\to(x,y)$ link is in ${\mathcal C}$ and oriented in the same direction, $-1$ if its orientated in the opposite direction, and $0$ if it is not in ${\mathcal C}$.
Then each current can be decomposed as 
\begin{equation}\label{eq:cycleSplit}
\begin{split}
{\mathcal J}_{x,x^\prime}^{y,y^\prime}=&\sum_{\mathcal C}\delta_{x,x^\prime}^{y,y^\prime}({\mathcal C}){\mathcal J}(\mathcal{C})+\sum_{\mathcal C_X}\delta_{x,x^\prime}^{y,y^\prime}({\mathcal C}_X){\mathcal J}(\mathcal{C}_X) \\
&+\sum_{\mathcal C_Y}\delta_{x,x^\prime}^{y,y^\prime}({\mathcal C}_Y){\mathcal J}(\mathcal{C}_Y)
\end{split}
\end{equation}
where we have separated out the sum on global fundamental cycles ${\mathcal C}$ from local ones, ${\mathcal C}_X$ and ${\mathcal C}_Y$.

We describe how this can be used to modify ${\dot S}^X_{\bf i}$, the same argument applies to ${\dot S}^Y_{\bf i}$. 
Upon substitution of Eq.~(\ref{eq:cycleSplit}) into Eq.~(\ref{eq:2lawSplitSS}), we find
\begin{align}
\dot{\mathcal S}_{\bf i}^X=&\sum_{x\ge x^\prime;y}{\mathcal J}_{x,x^\prime}^y\left(\mathcal{F}_{x,x^\prime}^y-f_{x,x^\prime}^y\right) \\
=&\sum_{\mathcal C} {\mathcal J}({\mathcal C})\sum_{x\ge x^\prime;y}\delta_{x,x^\prime}^{y}({\mathcal C})({\mathcal F}_{x,x^\prime}^y-f_{x,x^\prime}^y) \\
&+\sum_{\mathcal C_X}{\mathcal J}(\mathcal{C}_X)\sum_{x\ge x^\prime;y}\delta_{x,x^\prime}^{y}({\mathcal C}_X){\mathcal F}_{x,x^\prime}^y \nonumber,
\end{align}
after recognizing that there is no contribution to ${\dot S}^X_{\bf i}$ on local $Y$-cycles ${\mathcal C}_Y$ and that information only acts on global cycles, $\sum \delta_{x,x^\prime}^{y}({\mathcal C}_X)f_{x,x^\prime}^y=0$ for all ${\mathcal C}_X$.
We arrive at Eq.~\eqref{eq:cycleEnt} by identifying ${\mathcal F}^X({\mathcal C})=\sum\delta_{x,x^\prime}^{y}({\mathcal C}){\mathcal F}_{x,x^\prime}^y$ and ${\mathcal F}^I({\mathcal C})=\sum\delta_{x,x^\prime}^{y}({\mathcal C})f_{x,x^\prime}^y$.

\bibliography{PhysicsTexts,Feedback} 

\end{document}